\documentclass[%
 reprint,
superscriptaddress,
 amsmath,amssymb,
 aps,
pra,
floatfix
]{revtex4-1}
\usepackage{amssymb, blindtext, amsmath, lineno, placeins, graphicx, capt-of}
\renewcommand{\vec}[1]{\mathbf{#1}}
\usepackage{graphicx}
\usepackage{dcolumn}
\usepackage{bm}

\begin{document}

\preprint{APS/123-QED}

\title{Surface Plasmon Polaritons Sustained at the Interface of a Nonlocal Metamaterial}

\author{Joshua Feis}
\affiliation{Institute of Theoretical Solid State Physics, Karlsruhe Institute of Technology, Wolfgang-Gaede-Stra\ss e 1, 76131 Karlsruhe, Germany}
\author{Karim Mnasri}
\email{karim.mnasri@kit.edu}
\affiliation{Institute of Theoretical Solid State Physics, Karlsruhe Institute of Technology, Wolfgang-Gaede-Stra\ss e 1, 76131 Karlsruhe, Germany}
\affiliation{Institute of Applied and Numerical Mathematics, Karlsruhe Institute of Technology, Englerstrasse 2, 76131, Karlsruhe, Germany}
\author{Andrii Khrabustovskyi}
\affiliation{Institute of Applied Mathematics, Graz University
of Technology, Steyrergasse 30, 8010 Graz, Austria}
\author{Christian Stohrer}
\affiliation{Institute of Applied and Numerical Mathematics, Karlsruhe Institute of Technology, Englerstrasse 2, 76131, Karlsruhe, Germany}
\author{Michael Plum}
\affiliation{Institute for Analysis, Karlsruhe Institute of Technology, Englerstr. 2, 76131 Karlsruhe, Germany}
\author{Carsten Rockstuhl}
\affiliation{Institute of Theoretical Solid State Physics, Karlsruhe Institute of Technology, Wolfgang-Gaede-Stra\ss e 1, 76131 Karlsruhe, Germany}
\affiliation{Institute of Nanotechnology, Karlsruhe Institute of Technology, P.O. Box 3640, 76021 Karlsruhe, Germany}

\date{\today}

\begin{abstract}
Studying basic physical effects sustained in metamaterials characterized by specific constitutive relation is a research topic with a long standing tradition. Besides intellectual curiosity, it derives its importance from the ability to predict observable phenomena that are, if found with an actual metamaterial, a clear indication on its properties. Here, we consider a nonlocal (strong spatial dispersion), lossy, and isotropic metamaterial and study the impact of the nonlocality on the dispersion relation of surface plasmon polaritons sustained at an interface between vacuum and such metamaterial. For that, Fresnel coefficients are calculated and appropriate surface plasmon polaritons existence conditions are being proposed. Predictions regarding the experimentally observable reflection from a frustrated internal reflection geometry are being made. A different behavior for TE and TM polarization is observed. Our work unlocks novel opportunities to seek for traces of the nonlocality in experiments made with nowadays metamaterials. 
\begin{description}
\item[PACS numbers]
TBD
\end{description}
\end{abstract}

\maketitle

\section{Introduction}
Metamaterials are artificial optical materials that are made from, mostly, periodically arranged inclusions. They are studied out of scientific interest but also because they unlock in perspective unprecedented applications \cite{engheta2002ideas}. Among many examples, we mention perfect lenses \cite{GUENNEAU2009352}, invisibility cloaks \cite{PhysRevB.78.075107,monti2015optical}, and electromagnetic black holes \cite{narimanov2009optical}. One of the main problems in investigating metamaterials with complicated and densely packed geometries concerns their effective description \cite{PhysRevB.84.075153}. The goal of finding this description, known as homogenization, is to associate the actual metamaterial to a homogeneous material that has the same optical response. Optical response here means that once the material is homogenized, it can be considered in other geometries and optical settings and the description of the interaction of electromagnetic fields with this homogeneous material continues to be the same as if the actual metamaterial would have been considered. The process of homogenization can be considered as a two step process. First, a suitable constitutive relation is chosen that is expected to cover all emerging effects. Second, by choosing one among many possible technical means, the actual material parameters are retrieved. Frequently, a particular dispersion, i.e. a frequency dependency, can be assumed for the material properties. The functional dependency is motivated by basic phenomenological modeling \cite{PhysRevB.75.115104}.     

In the cases where the inclusions are only weakly interacting and much smaller than the excitation wavelength, local constitutive relations turned out to be fully sufficient \cite{PhysRevE.85.066603,PhysRevB.88.125131}. Then, the electric and magnetic field induce only at the considered spatial position an electric polarization or a magnetization, respectively. While considering for simplicity isotropic materials with no electro-magnetic coupling, the well-known Drude and/or Lorentz models for permeability and permittivity do frequently emerge for the frequency dependence of the material properties \cite{serdiukov2001electromagnetics}. This can be explained by considering on phenomenological grounds the inclusions to cause either a response associated to free electrons, e.g. in a straight wire elements \cite{PhysRevE.73.046612,zhao2006spatially}, or a response associated to a harmonic oscillator, e.g. in a small metallic or dielectric particle that is driven into a resonant optical response, in a split ring \cite{PhysRevE.72.026615,WANG2015491}, or any other complicated inclusion that has been suggested in the past \cite{DU2008730,PhysRevB.77.195328,Wu2007}.

However, when the metamaterials are operated at wavelengths that are not much longer than the size of the inclusion but only gently longer or the inclusions themselves show a strong coupling to their neighbors, a local description at the effective level fails to capture the properties of the metamaterial \cite{PhysRevB.81.035320}. An electric field at one point in space can then induce a response at a distant point in space. To improve the description, it is therefore necessary to go beyond the local description and take into account the nonlocality; that is also called a strong spatial dispersion. 

Recently, a model has been proposed \cite{PhysRevB.97.075439} introducing advanced nonlocal constitutive relations that read as
\begin{flalign}
\vec{D}(\vec{r},\omega)  = \boldsymbol\epsilon \vec{E}(\vec{r},\omega) &+ \nabla \times \boldsymbol\gamma \nabla\times \vec{E}(\vec{r},\omega) \label{strong_spatial_dispersion}
\\ &+  \nabla \times \nabla \times \boldsymbol{\eta}\nabla \times \nabla \times \vec{E}(\vec{r},\omega) \notag
\end{flalign}
and
\begin{equation}
\vec{B}(\vec{r},\omega)  = \vec{H}(\vec{r},\omega)
\end{equation}
with material parameters $\boldsymbol\epsilon$, $\boldsymbol\gamma$, and $\boldsymbol\eta$. For convenience, we assume that the coordinate system of the laboratory is aligned to the coordinate system of the principle axis of the metamaterial. Hence, all material parameters are considered to be diagonal tensors. By a suitable gauge transformation, $\boldsymbol\gamma$ can be recast to appear as a dispersion in the local permeability. For that reason it is called a weak spatial dispersion. In contrast, $\boldsymbol\eta$ is associated to a strong spatial dispersion, it is a nonlocal material parameter. These constitutive relations allow for a rigorous mathematical derivation of interface conditions that extend the known conditions from basic electrodynamics \cite{PhysRevB.97.075439}. Therefore, not just light propagation in bulk material can be described but also functional elements thereof. 

In this contribution, we continue a long standing tradition in metamaterials research in where basic physical effects are explored once a specific material with given properties has been assumed. On the one hand, such research is intellectual appealing, as novel effects that were predicted with those materials constitute a major driving force to finally identify materials that offer these properties. This may have started with the seminal work from Veselago \cite{veselago1968electrodynamics}. There, he simply has been assuming the availability of a material with a dispersive permittivity and a dispersive permeability and studied afterwards observable optical effects. More contemporary examples would be the field of transformation optics or the suggestion for a perfect lens \cite{chen2010transformation,0953-8984-15-37-004,Pendry03}. On the other hand, from such consideration we can predict experimentally observable features that would be a smoking gun to decide whether a particular constitutive relation indeed is necessary for the effective description of a metamaterial or not. 

Here, we investigate propagating surface plasmon polaritons (SPP) at the interface between an ordinary material and a nonlocal metamaterial that exhibits a negative refractive index in some frequency range of interest. A list of metamaterials that have been designed to undergo such properties can be found in \cite{Soukoulis47}. Our contribution is particularly inspired by the work of Ruppin \cite{Ruppin2000SurfaceMedium} that pioneered the study of surface waves sustained at the interface between metamaterials described by local constitutive relations and ordinary media. Specifically, we investigate the dispersion relation of surface waves and the reflection to be observed in a frustrated total internal reflection geometry and how these are affected by a strong nonlocality. With that, we offer blueprints for experiments that aim to elucidate nonlocal properties in metamaterials.

Our manuscript is structured such that we introduce in the next section the material models we consider. We study then the interface Fresnel equations, which need to be known in order to derive the dispersion relation of the propagating SPPs. In section four, we study the dispersion relation in the presence of nonlocality where we particular emphasize the question how the onset of a weak nonlocality changes the dispersion relation of the propagating SPP sustained at the interface between vacuum and the metamaterial. In a last step, experimental features are predicted as observable in an attenuated total reflection setup. Finally, we conclude on our research.

\section{Material Models and Dispersion Relations}

\begin{figure}[bt]
		\centering
		\includegraphics[scale=0.8]{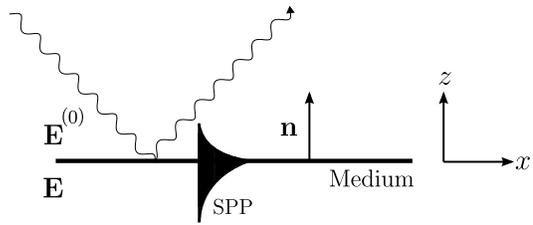}
		\caption[Schema of surface plasmon polaritons at an interface.]{Schema of surface polaritons sustained at an interface. The interface at $z=0$ divides two extended half-spaces. One is vacuum while the other one is made initially from an isotropic homogeneous medium with permittivity $\epsilon$ and permeability $\mu$. Later, also a more complicated constitutive relation is considered. The light is not transmitted but excites surface plasmon polaritons (SPP) on the interface. For this purpose, it is required to have a wave vector component in the direction of the interface that is larger than the length of the wave vector in vacuum.}
	\label{fig:SPPSkizze}
\end{figure}

In order to model a metamaterial with negative index behavior, we consider a homogeneous, isotropic, and nonlocal metamaterial with a Drude permittivity
\begin{equation}
\epsilon(\omega) = 1 - \frac{\omega_\mathrm{p}^2}{\omega(\omega+\mathrm{i}\Gamma_\epsilon)} \label{eq:Drude}
\end{equation}
and a permeability described by a Lorentz model 
\begin{equation}
\mu(\omega) = 1-\frac{F\omega^2}{\omega^2-\omega_0^2+\mathrm{i}\Gamma_\mu \omega}\,, \label{eq:Lorentz}
\end{equation}
with parameters $\omega_p$, $\omega_0$, $\Gamma_{\epsilon}$, $\Gamma_{\mu}$ and $F$, which will be commented on below.
Via a gauge transformation
\begin{equation}
\vec{D'}(\vec{r},\omega) = \vec{D}(\vec{r},\omega) + \nabla \times \vec{Q}(\vec{r},\omega) 
\end{equation}
and
\begin{equation}
\vec{H'}(\vec{r},\omega) = \vec{H} - \mathrm{i} k_0 \vec{Q}(\vec{r},\omega), 
\end{equation}
the parameter $\gamma(\omega)$ that is associated to a weak spatial dispersion can be expressed through $\mu(\omega)$:
\begin{equation}
\gamma(\omega) = \frac{\mu(\omega) - 1}{ \mu(\omega) k_0^2} \label{eq:gamma},
\end{equation}
with $k_0=\omega/c$, by choosing $\vec{Q}(\vec{r},\omega) \overset{!}{=} - \gamma \nabla \times \vec{E}(\vec{r},\omega)$. 

For the parameter expressing the strong spatial dispersion we make a model assumption and set its frequency dependency to
\begin{equation}
\eta(\omega) = \frac{GF\omega^2}{\omega_0^2-\omega^2-\mathrm{i}\Gamma_\eta \omega}\,, \label{eq:ETA}
\end{equation}
where $G$ is a constant parameter scaling the strength of the nonlocality. With that we tight the nonlocality to the magnetic response. This seems to be meaningful as in the resonance an enhanced light-matter-interaction can be observed that will cause the nonlocal response. The resonance frequency of the material is chosen to be {$\omega_0 = 4$ GHz}, the plasma frequency to be $\omega_\mathrm{p} = 10$ GHz. Further, we set $F = 0.56$ and the damping terms shall be {$\Gamma_\epsilon = 0.03\, \omega_\mathrm{p}$} and $\Gamma_\mu = \Gamma_\eta = 0.03\, \omega_0$. For the local material parameters, the same values were used by Ruppin in \cite{Ruppin2000SurfaceMedium}. The exact values are of no importance and only serve the purpose to make everything definite.

We restrict ourselves to a plane geometry where the half-space below the $x$-axis is filled with the metamaterial and the half-space above it with vacuum, as illustrated in Fig.~\ref{fig:SPPSkizze}. Because the metamaterial is modeled as a homogeneous material and the vacuum is trivial, the eigenmodes and hence the fields can be described by plane waves
\begin{equation}
\vec{E}^{(i)}(\vec{r},\omega) \propto \exp[\mathrm{i}(hx + k_z^{(i)} z)  ] .
\end{equation}
$\vec{E}^{(0)}(\vec{r},\omega)$ shall be the field in the vacuum with 
\begin{equation}
(k_z^{(0)})^2=k_0^2-h^2 \label{eq:VDR}
\end{equation}
being the coresponding dispersion relation. For the metamaterial with a strong spatial dispersion, characterized by the constitutive relation given in Eq.~\eqref{strong_spatial_dispersion}, the dispersion relation reads as
\begin{equation}
(k_z^{(i)})^2 = -h^2 + p \pm \sqrt{p^2 - q} \label{eq:NLDR}
\end{equation}
with $q = \epsilon/\eta$ and $p = (2k_0^2 \eta \mu)^{-1}$ \cite{PhysRevB.97.075439}. 

This relation has four mathematical solutions in total, two of which are physical. We denote these two with the indexes $i=1$ and $i=2$. Note, that they only differ in the sign in front of the square root. The others result in waves diverging towards infinity due to their positive imaginary part. In the lossless case, for the existence of SPPs, it is usually required that the wave vector components away from the interface are purely imaginary, so that the wave is evanescent in that direction. However, in the more realistic case of an absorptive medium as considered here, this condition is not feasible \cite{warmbier2012surface}. Instead, we only require the radiation away from the interface to be very strongly dampened, so 
\begin{equation}
|\Im(k_z^{(i)} )| > |\Re(k_z^{(i)} )|. \label{eq:SPPCON}
\end{equation}
Using this condition, SPP with small radiative losses can also be discussed.

\section{Fresnel Coefficients}
As mentioned above, interface conditions have been derived for the model of nonlocality discussed here. Going on from these, the corresponding Fresnel equations have also been found \cite{PhysRevB.97.075439}. They read
\begin{equation}
\vec{F}_\mathrm{TM}\begin{pmatrix}
E_z^\mathrm{R} \\ E_z^{(1)} \\E_z^{(2)}
\end{pmatrix} = - E_z^\mathrm{I} 
\begin{pmatrix}
 k_z^\mathrm{I} \\ 1 \\0 \label{eq:FTM}
\end{pmatrix}
\end{equation}
for TM polarization with the TM Fresnel matrix 
\begin{equation}
\vec{F}_\mathrm{TM} \equiv
\begin{pmatrix}
k_z^\mathrm{R} & -k_z^{(1)} &-k_z^{(2)} \\ 
1 & - \epsilon  & - \epsilon \\
0&\eta  k_z^{(1)}  \left|\vec{k}^{(1)}\right|^2 & \eta  k_z^{(2)}  \left|\vec{k}^{(2)}\right|^2
\end{pmatrix},
\end{equation}
where $\left|\vec{k}^{(i)}\right|^2 = (k_z^{(i)})^2 + h^2$, and
\begin{equation}
\vec{F}_\mathrm{TE}\begin{pmatrix}
E_x^\mathrm{R} \\ E_x^{(1)} \\E_x^{(2)}
\end{pmatrix} = - E_x^\mathrm{I} 
\begin{pmatrix}
1 \\ k_z^\mathrm{I} \\0 \label{eq:FTE}
\end{pmatrix}
\end{equation}
for TE polarization, where we introduced the TE Fresnel matrix
\begin{equation}
\vec{F}_\mathrm{TE} \equiv \begin{pmatrix}
1 & -1 & -1 \\ 
k_z^\mathrm{R} & k_z^{(1)} A_1 &k_z^{(2)} A_2 \\
0 &\eta k_z^{(1)}& \eta k_z^{(2)}
\end{pmatrix}
\end{equation}
with $A_i = \left[-\mu^{-1} + \eta k_0^2\left|\vec{k}^{(i)}\right|^2\right]$. Using these equations, the reflection coefficients
\begin{widetext}
\begin{equation}
r^\mathrm{TM} = \frac{\epsilon k_z^\mathrm{I}\left[h^2 + (k_z^{(1)})^2+(k_z^{(2)})^2+ k_z^{(1)} k_z^{(2)}\right]-k_z^{(1)}k_z^{(2)}(k_z^{(1)}+k_z^{(2)})}{\epsilon k_z^\mathrm{R}\left[h^2 + (k_z^{(1)})^2+(k_z^{(2)})^2+ k_z^{(1)} k_z^{(2)}\right]-k_z^{(1)}k_z^{(2)}(k_z^{(1)}+k_z^{(2)})} \label{eq:RTM}
\end{equation}
for TM polarization and
\begin{equation}
r^\mathrm{TE} = \frac{\mu k_z^\mathrm{I}(k_z^{(1)}+k_z^{(2)}) + h^2 -k_z^{(1)}k_z^{(2)} - \eta \mu k_0^2 \left|\vec{k}^{(1)}\right|^2 \left|\vec{k}^{(2)}\right|^2}{\mu k_z^\mathrm{R}(k_z^{(1)}+k_z^{(2)}) + h^2-k_z^{(1)}k_z^{(2)} - \eta \mu k_0^2\left|\vec{k}^{(1)}\right|^2  \left|\vec{k}^{(2)}\right|^2} \label{eq:RTE}
\end{equation}
\end{widetext}
for TE polarization have been calculated. The reflection coefficients of a local medium can be reproduced from this by performing the limit $\eta \rightarrow 0$. One of the $k_z^{(i)}$ is divergent in this limit, without restriction let it be $k_z^{(2)}$.
Now, using $\lim_{\eta \rightarrow 0} k_z^{(i)} \eta = 0$,  $\lim_{\eta \rightarrow 0} (k_z^{(2)})^2 \eta = (k_0^2 \mu)^{-1}$ and $\lim_{\eta \rightarrow 0} (k_z^{(1)})^2 \eta = 0$ leads to the correct local limit for both reflection coefficients,
\begin{equation}
\lim_{\eta\rightarrow 0}r^\mathrm{TM} = \frac{\epsilon(\omega) k_z^\mathrm{I}-\lim_{\eta\rightarrow 0}k_z^{(1)}}{\epsilon(\omega)k_z^\mathrm{R}-\lim_{\eta\rightarrow 0}k_z^{(1)}}
\end{equation}
and
\begin{equation}
\lim_{\eta\rightarrow 0}r^\mathrm{TE} =\frac{k_z^\mathrm{I}-\mu(\omega)\lim_{\eta\rightarrow 0}k_z^{(1)}}{k_z^\mathrm{R}-\mu(\omega)\lim_{\eta\rightarrow 0}k_z^{(1)}} .
\end{equation}
\section{SPP Dispersion with Nonlocality}
\begin{figure*}[tb]
		\centering
		\includegraphics[scale=0.8]{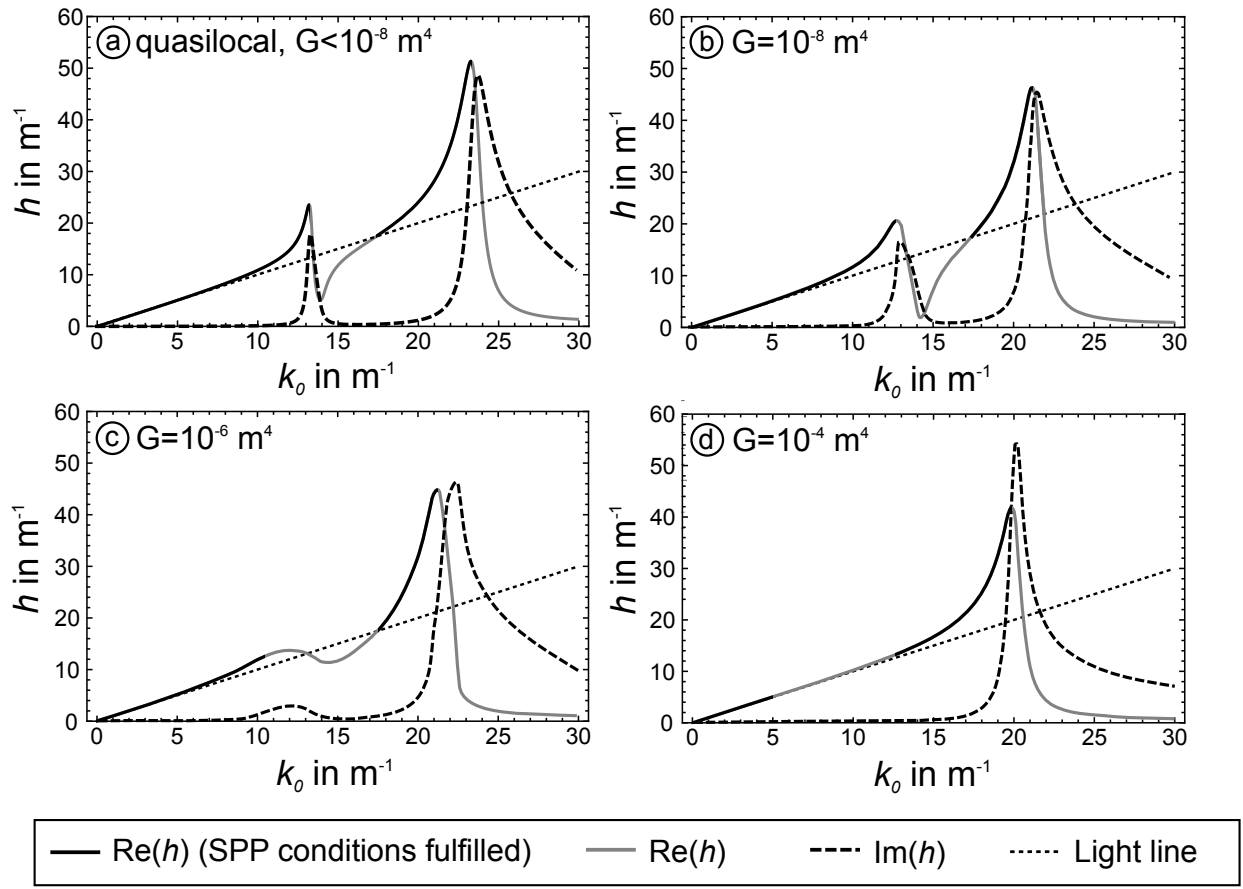}
		\caption[SPP Dispersion.]{SPP dispersion curve for TM polarized light.}
	\label{fig:DRTM}
\end{figure*}

\begin{figure*}[tb]
		\centering
		\includegraphics[scale=0.8]{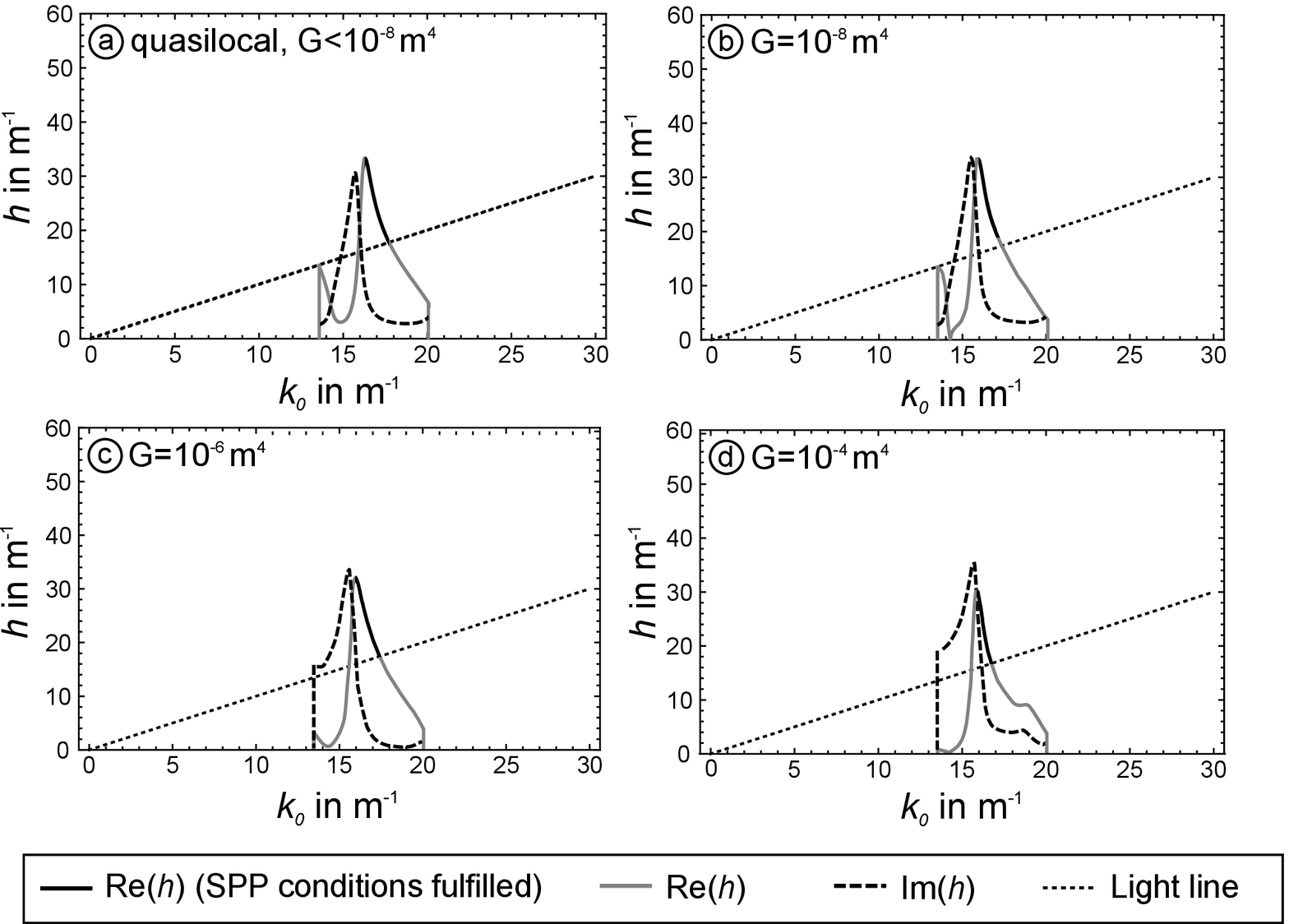}
		\caption[SPP Dispersion.]{SPP dispersion curve for TE polarized light.}
	\label{fig:DRTE}
\end{figure*}

The SPP dispersion relation can be found from the poles of the reflection coefficients in Eqs. \eqref{eq:RTE} and \eqref{eq:RTM} \cite{barchiesi2012classroom} with $k_z^\mathrm{R} = - k_z^\mathrm{I} = -k_z^{(0)}$. Making use of the simplifications
\begin{equation}
(k_z^{(1)}k_z^{(2)})^2 = h^4 + \frac{\epsilon}{\eta} -\frac{h^2}{k_0^2 \eta \mu} \label{eq:id1}
\end{equation}
and
\begin{equation}
(k_z^{(1)})^2 + (k_z^{(2)})^2 =  -2h^2 + ( k_0^2 \eta \mu)^{-1}, \label{eq:id2}
\end{equation}
these equations are solved. The solution formulas are very long and not displayed here for readability. Instead, numerical values as given above have been used to study the effect of the nonlocality on the SPP dispersion in Figs. \ref{fig:DRTM} and \ref{fig:DRTE}. 

Inequality \eqref{eq:SPPCON} gives an existence condition for SPP. This however, is not entirely sufficient. Calculating the ratio of the energy transmitted by each of the waves from the Fresnel equations \eqref{eq:FTE} and \eqref{eq:FTM}, 
\begin{equation}
\tau_\mathrm{TM} = \left|\frac{t_\mathrm{TM}^{(2)}}{t_\mathrm{TM}^{(1)}}\right|^2 =\left|\frac{k_z^{(1)}\left[h^2 + (k_z^{(1)})^2\right]}{k_z^{(2)}\left[h^2 + (k_z^{(2)})^2\right]}\right|^2
\end{equation}
for TM polarization and
\begin{equation}
\tau_\mathrm{TE} = \left|\frac{t_\mathrm{TE}^{(2)}}{t_\mathrm{TE}^{(1)}}\right|^2=\left|\frac{h^2 + (k_z^{(1)})^2}{h^2 + (k_z^{(2)})^2}\right|^2
\end{equation}
for TE polarization, we see that if one of the $k_z^{(i)}$ is very large compared to the other one, the associated wave carries a lot less energy. We require $0.01 < \tau_\mathrm{TE/TM} < 100$ as an additional condition, such that both waves carry at least 1\% of the energy. If this additional condition is not met, the original condition in Eq. \eqref{eq:SPPCON} is waived for the very large $k_z^{(i)}$. This needs to be done in order to ensure that even for divergent wave numbers, as they occur in the local limit, the SPP conditions proposed here remain meaningful. Finally, due to causality, only points outside the light cone are relevant, i.e. requiring $h>k_0$.
The resulting dispersion curves are displayed in Fig.~\ref{fig:DRTM} for TM polarization and in Fig.~\ref{fig:DRTE} for TE polarization.

Turning to the TM polarization first, the quasilocal case shown in Fig.~\ref{fig:DRTM} (a) exhibits a dispersion relation similar to the one obtained by Ruppin in \cite{Ruppin2000SurfaceMedium} but for a lossy medium. Quasilocal means that this curve can either be obtained by a very low nonlocality or the classical derivation going from the purely local wave equation and interface conditions. Instead of two divergent branches in the lossless case, it is now one connected curve that exhibits back-bending in place of the divergence. Opposed to the lossless case, not all solutions outside the light cone fulfill the SPP conditions. Scaling up to a low nonlocality with $G=10^{-8}\,\mathrm{m}^4$ [see Fig.~\ref{fig:DRTM} (b)], the dispersion relation stays similar to the quasilocal case, although a slight broadening with a decrease in height can be observed for the peaks in the dispersion relation. Again, SPP exist on the upward part of the peaks. Going to a moderate magnitude of the nonlocality with $G=10^{-6}\,\mathrm{m}^4$ [see Fig.~\ref{fig:DRTM} (c)], the first peak in the area of the resonance frequency flattens out much further. The higher frequency peak roughly stays the same. The frequency range in which the SPP conditions are fulfilled also doesn't change much for the higher frequencies while it shrinks for lower frequencies. Finally, the highest nonlocality amplitude considered is $G=10^{-4}\,\mathrm{m}^4$ due to the effect being of fourth order. In fact, even higher nonlocalities don't alter the shape of the dispersion curve any further. As can be seen in Fig.~\ref{fig:DRTM} (d), the lower peak has entirely vanished and now follows the light line. The upper peak didn't change its shape much again. SPPs can now exist in an even smaller frequency range for lower frequencies or on the upward slope of the peak. All in all, the nonlocality has led to a vanishing lower frequency peak while it didn't alter the higher frequency peak much. Looking at the chosen form of the nonlocality, we can conclude that in the region where the nonlocality has its maximum, for high nonlocalities the medium becomes non-dispersive along the surface and SPP cannot exist there anymore.

For TE polarization, the dispersion curves are displayed in Fig.~\ref{fig:DRTE}. There are only valid solutions to the dispersion relation in the frequency
range where the permeability takes on negative values. Similar to the TM dispersion relation, a comparison with the lossless case discussed by Ruppin \cite{Ruppin2000SurfaceMedium} shows that instead of the sharp divergence of the dispersion curve there is a back bending into the peak as shown in Fig.~\ref{fig:DRTE} (a). SPPs can exist on the downward slope of that peak. Figure~\ref{fig:DRTE} (b) shows the change when taking into account a small nonlocality with $G=10^{-8}\,\mathrm{m}^4$. There is a slight change in the shape of the dispersion curve, but the part that lies outside the light cone stays almost the same, just as the frequency range in that the SPP conditions are fulfilled. Going to a moderate nonlocality of $G=10^{-6}\,\mathrm{m}^4$, the shape again stays almost the same apart from a slight decrease in the peak maximum. Just as previously, SPPs exist on the downward slope of the peak outside the light cone. Finally, the highest nonlocality of $G=10^{-4}\,\mathrm{m}^4$ decreases the peak height further and leads to a very slightly narrower peak but only significantly changes the shape of the dispersion curve below the light line. In summary, it can be said that this type of nonlocality does not affect the TE mode SPP a lot and opposed to the TM mode, where the changes are quite significant, only causes slight changes in the dispersion relation. 

\section{ATR Spectra}

Surface plasmon polariton excitation can be observed using the method of attenuated total reflection (ATR) spectroscopy. A geometry for that was proposed by Otto \cite{otto1968excitation}. 

\begin{figure}[tb]
		\centering
		\includegraphics[scale=0.8]{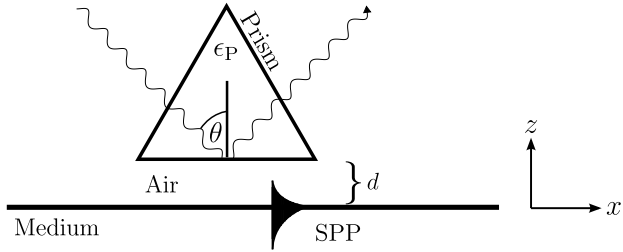}
		\caption[]{Schema of the Otto configuration for attenuated total reflection (ATR) spectroscopy. The angle of incidence $\theta$ is chosen to be the angle of total internal reflection resulting in an evanescent wave in the air layer, that couples to the SPPs in the medium at the air-metamaterial interface, where $z=0$.}
	\label{fig:OTTO}
\end{figure}

It consists of a prism with permittivity $\epsilon_\mathrm{P}$, an air layer of thickness $d$, and the medium to be analyzed, so in this case the metamaterial. Light is sent in at an angle $\theta$ that it is totally reflected at the prism-air interface. The evanescent waves penetrating the air layer can then excite SPPs at the air-metamaterial interface. This geometry, commonly referred to as the Otto configuration, is illustrated in Fig.~\ref{fig:OTTO}. Using a transfer matrix formalism \cite{ohta1990matrix} and the previously derived reflection coefficients in Eq. \eqref{eq:RTM} and Eq. \eqref{eq:RTE}, the reflectivity of such a setup has been calculated for both polarizations. For the metamaterial, the same parameters as above have been used. The permittivity of the prism is chosen to be nondispersive with $\epsilon_\mathrm{P}=3$.

\begin{figure*}[t]
\centering
\begin{minipage}{.49\textwidth}
  \centering
  \includegraphics[scale=0.55]{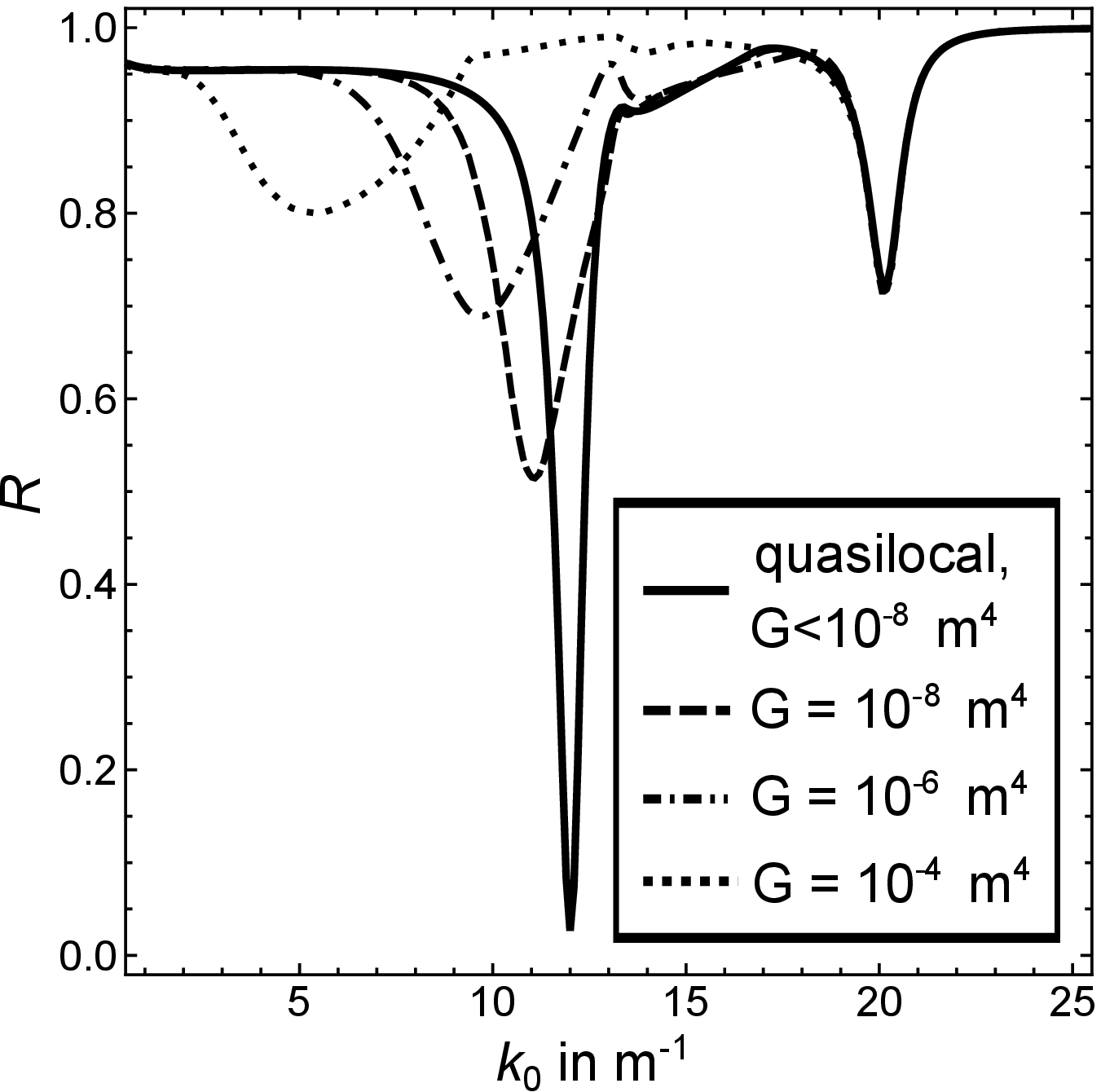}
  \captionof{figure}{Reflection coefficient of TM polarized light calculated for Otto configuration with $d=3$ cm and $\theta = \pi/4$.}
  \label{fig:ATRTM}
\end{minipage}%
\hspace{0.2cm}
\begin{minipage}{.49\textwidth}
  \centering
  \includegraphics[scale=0.55]{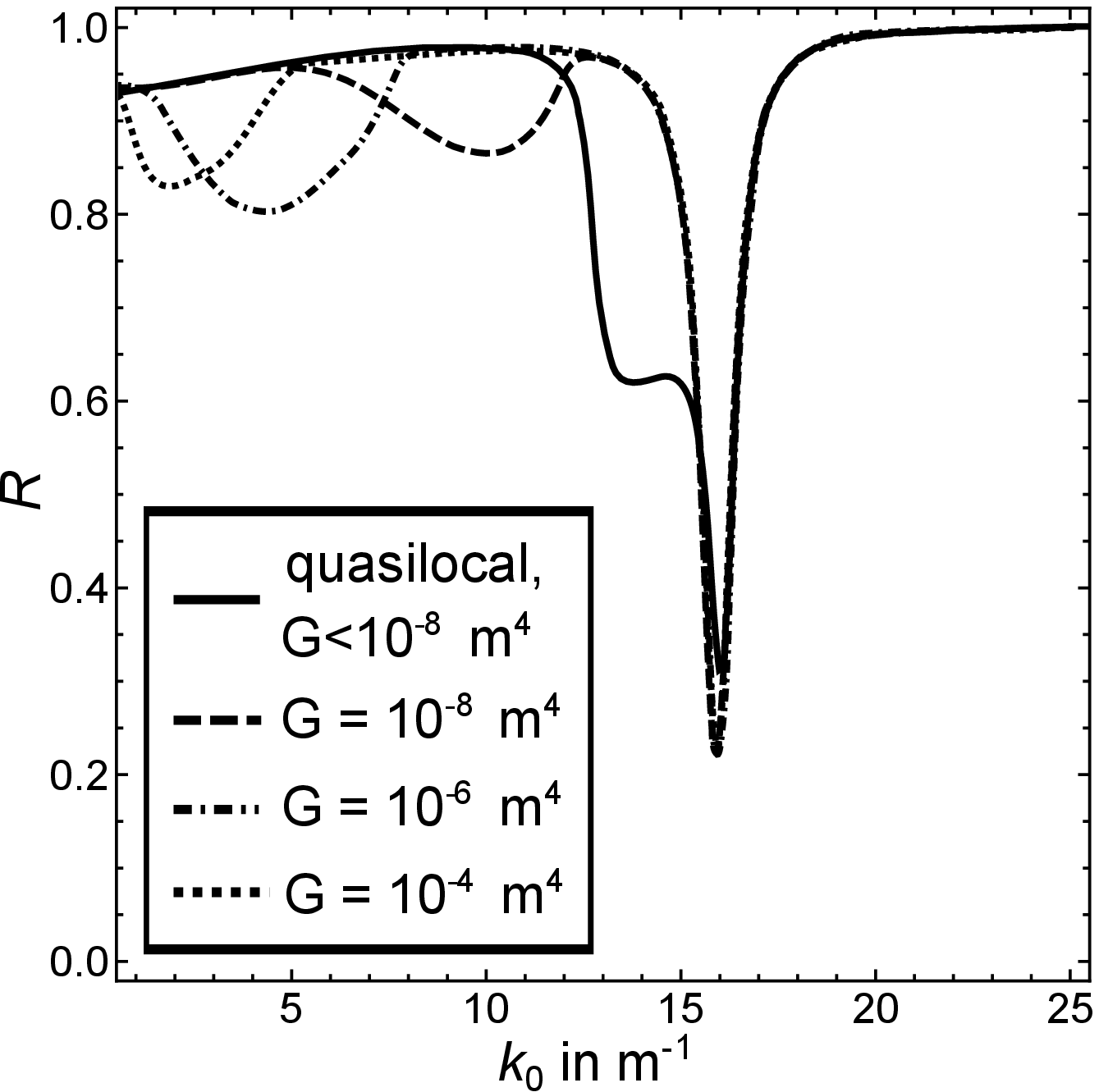}
  \captionof{figure}{Reflection coefficient of TE polarized light calculated for Otto configuration with $d=1$ cm and $\theta = \pi/3$.}
  \label{fig:ATRTE}
\end{minipage}
\end{figure*}

Turning to TM polarization first, with an angle of incidence of $\theta = \pi/4$ and thickness $d=3\,\mathrm{cm}$, the spectra in Fig.~\ref{fig:ATRTM} have been obtained. The quasilocal curve has two SPP peaks, one for each SPP branch, and is identical to the one predicted by Ruppin \cite{Ruppin2000SurfaceMedium}. For all different magnitudes of the nonlocality, the higher frequency peak stays almost unchanged. Belonging to the higher frequency peak in the SPP dispersion relation (Fig.~\ref{fig:DRTM}), this observation confirms the prediction that there is no significant change to that peak. The lower frequency peak in the spectrum however, gets less pronounced and drifts off to lower frequencies. This is associated with shrinking of the frequency range that the SPP conditions are fulfilled for with increasing nonlocality. Correspondingly, the frequencies that SPP can be excited are lower and lower.
Additionally, the decreasing depth of the peak implies that the coupling strength to the SPP mode gets less with an increase in nonlocality. The very small peak seen inbetween the two SPP peaks is due to the excitation of bulk polaritons \cite{Ruppin2000SurfaceMedium}. With an increase in nonlocality, this peak also becomes less pronounced. 

The spectrum for TE polarization is shown in Fig.~\ref{fig:ATRTE}. The parameters used here are $\theta = \pi/3$ for the angle of incidence and $d=1\,\mathrm{cm}$ for the distance between the prism and the metamaterial. Again, the quasilocal curve is identical to Ruppin's work. The minimum at higher frequencies of that curve is due to the excitation of SPP while the one at lower frequencies is due  to the excitation of bulk polaritons. An increase of the nonlocality leads to the bulk polariton peak moving to lower frequencies and becoming less deep. The SPP peak keeps its form and position independent of the nonlocality. This is in compliance with the dispersion relation (Fig.~\ref{fig:DRTE}), which predicted no significant effect despite the increasing nonlocality.

\section{Conclusions}

Concluding, we have discussed the Fresnel equations for an interface between a nonlocal, homogeneous and isotropic metamaterial and vacuum and derived expressions  for the reflection coefficients for both TE and TM polarizations. Further, we proposed appropriate existence conditions for SPP in lossy, nonlocal media. Using these results, we obtained the dispersion relation for surface plasmon polaritons and discussed the effect of a gradually increasing nonlocality on it. We observed that the nonlocality has no significant effect in the case of TE polarized light. For TM polarized light however, it leads to the collapse of the lower frequency dispersion peak to a non-dispersive form while the higher frequency peak does not change its shape significantly. These observations were backed up by calculating and discussing the ATR spectrum in an Otto configuration.

\section*{Acknowledgements}
We gratefully acknowledge financial support by the Deutsche Forschungsgemeinschaft (DFG) through CRC 1173. K.M. also acknowledges support from the Karlsruhe School of Optics and Photonics (KSOP). C.S. acknowledges the support of the Klaus Tschira Stiftung.

\bibliography{bibliography}

\end{document}